%% file: paper.tex
\documentclass[sigconf, nonacm]{acmart}

\usepackage{glossaries}                 
\usepackage{amsfonts}                   
\usepackage{pifont}                     
\usepackage{color-edits}                
\usepackage{xcolor}                     
\usepackage{enumitem}                   
\usepackage{subcaption}                 
\usepackage{graphicx}                   
\usepackage{cleveref}                   
\usepackage{makecell}                   
\usepackage{booktabs}                   
\usepackage{multirow}                   
\usepackage{xfrac}                      
\usepackage{float}                      
\usepackage{dblfloatfix}                
\usepackage{cleveref}                   

\crefname{algocf}{Alg.}{Algs.}
\crefname{figure}{Fig.}{Figs.}
\crefname{equation}{Eq.}{Eqs.}

\setacronymstyle{long-short}
\newacronym{sram}{SRAM}{static random access memory}
\newacronym{ici}{ICI}{inter-chiplet interconnect}
\newacronym{icis}{ICIs}{inter-chiplet interconnects}
\newacronym{dse}{DSE}{design space exploration}
\newacronym{phys}{PHYs}{physical layers}
\newacronym{d2d}{D2D}{die-to-die}
\newacronym{c4}{C4}{controlled collapse chip connection}
\newacronym{noc}{NoC}{network-on-chip}
\newacronym{c2c}{C2C}{compute-to-compute}
\newacronym{c2m}{C2M}{compute-to-memory}
\newacronym{c2i}{C2I}{compute-to-IO}
\newacronym{m2i}{M2I}{memory-to-IO}
\newacronym{spm}{SPM}{scratchpad memory}
\newacronym{re}{RE}{recurring engineering}
\newacronym{nre}{NRE}{non-recurring engineering}
\newacronym{hbm}{HBM}{high-bandwidth memory}
\newacronym{shg}{SHG}{sparse Hamming graph}
\newacronym{fb}{FB}{flattened butterfly}

\let\mathpi\pi
\renewcommand{\pi}[1]{\textbf{\mathpi{#1}}}

\newif\ifnb     
\nbfalse
\nbtrue

\newif\ifcom    
\comfalse

\newif\ifps     
\psfalse

\addauthor[Summary]{ps}{orange}
\addauthor[Patrick]{pi}{teal}

\renewcommand{\pi}[1]{\ifcom\picomment{#1}\fi}

\setlist{leftmargin=1em}

\begin{document}

\title{RapidChiplet: A Toolchain for Rapid Design Space Exploration of Chiplet Architectures}

\ifnb
	\author{Patrick Iff}
	\email{patrick.iff@inf.ethz.ch}
	\orcid{0000-0001-5979-4915}
	\affiliation{%
	  \institution{ETH Zurich}
	  \city{Zurich}
	  \country{Switzerland}
	}

	\author{Benigna Bruggmann}
	\email{benigna.bruggmann@alumni.ethz.ch}
	\orcid{0000-0002-5730-0501} 
	\affiliation{%
	  \institution{ETH Zurich}
	  \city{Zurich}
	  \country{Switzerland}
	}

	\author{Blaise Morel}
	\email{blaise.morel@inf.ethz.ch}
	\affiliation{%
	  \institution{ETH Zurich}
	  \city{Zurich}
	  \country{Switzerland}
	}

	\author{Maciej Besta}
	\email{maciej.besta@inf.ethz.ch}
	\orcid{0000-0002-6550-7916} 
	\affiliation{%
	  \institution{ETH Zurich}              
	  \city{Zurich}
	  \country{Switzerland}
	}

	\author{Luca Benini}
	\email{lbenini@iis.ee.ethz.ch}
	\orcid{0000-0001-8068-3806} 
	\affiliation{%
	  \institution{ETH Zurich, Zurich, Switzerland}
	  \country{University of Bologna, Bologna, Italy}
	}

	\author{Torsten Hoefler}
	\email{torsten.hoefler@inf.ethz.ch}
	\orcid{0000-0002-1333-9797} 
	\affiliation{%
	  \institution{ETH Zurich}
	  \city{Zurich}
	  \country{Switzerland}
	}
\fi

\renewcommand{\shortauthors}{Iff et al.}

\begin{abstract}
	\input{abstract}

\end{abstract}

\maketitle

\begin{center}
\ifnb
\vspace{-0.5em}
\textbf{Website \& code:} https://github.com/spcl/rapidchiplet
\vspace{-0.5em}
\fi
\end{center}

\input{01_introduction}
\input{02_rapid_chiplet}

\input{03_evaluation}

\input{04_case_study}

\input{05_related_work}

\input{06_conclusion}

\ifnb
\input{07_acknowledgements}
\fi

\bibliographystyle{IEEEtran}
\bibliography{bibliography}

\end{document}

%% file: abstract.tex
Chiplet architectures are on the rise as they promise to overcome the scaling challenges of monolithic chips.
A key component of such architectures is an efficient inter-chiplet interconnect (ICI).
The ICI design space is huge as there are many degrees of freedom such as the number, size, and placement of chiplets, the topology and bandwidth of links, the packaging technology, and many more.
While ICI simulators are important to get reliable performance estimates, they are not fast enough to explore hundreds of thousands of design points or to be used as a cost function for optimization algorithms or machine learning models.
To address this issue, we present RapidChiplet, a fast and easy to use ICI latency and throughput prediction toolchain.
Compared to cycle-level simulations, we trade $0.25\%$--$30.15\%$ of accuracy for $427\times$--$137$,$682\times$ speedup.

%% file: 01_introduction.tex
\section{Introduction}
\label{sec:intro}

Increasing the performance-per-cost of processors and accelerators has become more challenging than ever, due to a slow-down of Moore's law \cite{mooreslaw}.
This slow-down is caused by the exponentially growing design and manufacturing costs when transitioning to a more advanced technology node \cite{cost-tech-node} as well as by the diminishing return of this transition due to the scaling limits of IO-drivers, analog circuits, and, most recently, \gls{sram}.
A promising mitigation strategy for these challenges leverages 2.5D stacking, where multiple silicon dies, called chiplets, are integrated into the same package.
The fact that a single chiplet design can be reused for multiple products helps to amortize the non-recurring costs.
Furthermore, since 2.5D stacking allows integrating heterogeneous chiplets that are built in different technology nodes into the same package, only components that can take full advantage of technology scaling are manufactured in advanced and costly technology nodes.
Components that have reached their scaling limits are manufactured in mature, lower-cost technologies.
These advantages have driven industry-leading companies to adopt 2.5D stacking for their products, e.g., Intel's Ponte Vecchio CPU \cite{ponte-vecchio}, AMD's EPYC and Ryzen CPUs \cite{amd-chiplets}, NVIDIA's Blackwell GPU \cite{blog_blackwell}, or Tesla's DOJO AI chip \cite{dojo}.

A crucial component of a 2.5D stacked chip is an \gls{ici} to provide connectivity between chiplets.
The \gls{ici}'s latency and throughput are affected by a wide variety of design choices, such as the packaging technology \cite{psi, intact, emib, dbhi}, number and size of chiplets \cite{chiplets-how-small}, chiplet placement \cite{hexamesh}, \gls{d2d} link implementation \cite{d2d-link-1,d2d-link-2}, communication protocol \cite{ucie, bow}, \gls{ici} topology \cite{double-butterfly, butterdonut, cluscross, kite, sid-mesh}, and many more.
Exploring this huge design space with the tools currently available is challenging due to two major reasons:
Firstly, existing tools for \gls{ici} latency and throughput estimation \cite{booksim, noxim, nostrum, garnet} usually rely on cycle-based timing simulations, which are time-consuming and hence only allow the exploration of a limited number of designs. 
Secondly, these tool often fail to capture the complex interplay between different \gls{ici} design decisions. 
For example, choosing an \gls{ici} topology with a higher router radix requires more \gls{phys} per chiplet, which increases the chiplet area, which in turn increases the link lengths, which has a packaging-technology-dependent effect onto the \gls{ici} latency and throughput.

To address these challenges, we introduce RapidChiplet, a fast and comprehensive toolchain for rapid \gls{dse} of \gls{ici}s.
RapidChiplet features high-level latency and throughput proxies that trade-off accuracy for speed, e.g., for transpose traffic, we achieve a speedup of $1$,$686\times$ (latency) and $102$,$110\times$ (throughput) compared to cycle-based timing simulations, at the cost of a $0.28\%$ (latency) and $17.89\%$ (throughput) deviation from the simulation results.
These proxies are intended for large-scale \gls{dse} where hundreds of thousands of designs need to be evaluated, or as a cost-function for optimization algorithms and machine learning models. 
To thoroughly evaluate a smaller number of selected designs, RapidChiplet provides a seamless integration of the cycle-based BookSim2 \cite{booksim} network simulator.
Furthermore, to capture the complex interplay between different \gls{ici} design decisions, RapidChiplet incorporates a wide range of configurable parameters, including, but not limited to, chiplet size, PHY-placement, \gls{ici} topology, packaging technology, chiplet placement, and routing algorithm.
To automate the \gls{dse}, we provide a rich toolbox that sweeps over user-defined parameter ranges, automatically generates the required inputs, runs RapidChiplet, and visualizes the results.

%% file: 02_rapid_chiplet.tex
\input{fig_overview}

\input{fig_inputs}

\section{The RapidChiplet Toolchain}
\label{sec:rc}

\Cref{fig:overview} outlines the RapidChiplet toolchain. 
The heart of the toolchain is the RapidChiplet core, which reads a wide range of user-defined inputs, validates them, and estimates the \gls{ici}'s latency and throughput.
In addition, we offer cycle-based flit-level simulations using BookSim2 \cite{booksim}.
Our toolchain automates the \gls{dse} s.t. the user only needs to specify parameter ranges (e.g., a list of topologies, chiplet-counts, packaging technologies, etc.) in an \textit{experiments} file.
The toolchain then iterates over all possible combinations of these parameters, generates the corresponding input files, and uses them to predict the \gls{ici}'s latency and throughput.
One of the few inputs that are not automatically generated are the traffic traces, which can be used for the cycle-based simulations in BookSim2.
We provide functionalities to export network traces from the Netraces v1.0 trace collection \cite{netraces} using Netrace \cite{netrace,netrace-report}, and to parse them into the RapidChiplet format.
The user can write custom parsing functions for arbitrary trace sources.
Finally, we provide visualization functions for results and inputs.

\subsection{RapidChiplet Core}
\label{ssec:rc-core}

\subsubsection{Inputs and Validation}
\label{sssec:rc-core-inputs}

\Cref{fig:inputs} visualizes the different inputs required by RapidChiplet.
Each input directory can contain multiple input files, and to execute RapidChiplet, one file from each directory is required.
To facilitate the handling of input files, a \textit{design} file is used to specify the paths to the remaining input files.
Input files can be combined at will as long as they are compatible (e.g., all chiplets used in the \textit{placement} must exist in the \textit{chiplet} file or chiplet-IDs in the \textit{routing table}, \textit{traffic}, or \textit{trace} files must exist in the \textit{placement} file). 
Our toolchain validates the inputs prior to each run.

\subsubsection{Latency Proxy}
\label{sssec:rc-core-latency}

We represent the \gls{ici} as an undirected, weight-ed graph $G = (V,E)$, where chiplets and on-interposer routers (for active silicon interposers only) are represented by vertices and the links between them are represented by edges. 
The weight $w(v)$ of a chiplet-vertex $v \in V$ is set to the chiplet's internal latency, and that of a router-vertex $v \in V$ is set to the router-latency.
The weight $w(\{u,v\})$ of an edge $\{u,v\} \in E$ is set to the link latency, which can be constant or link-length dependent.
In case of length-dependent link latencies, RapidChiplet computes all link-lengths, considering the chiplet positions and rotations, the placement of PHYs within the chiplets, and the link routing method (e.g., Manhattan, or direct).
If the link is connected to a chiplet rather than an on-interposer routers, the PHY-latency is added to $w(\{u,v\})$.
To estimate the average packet latency $L$ under a given \textit{traffic} $\mathcal{T}$, we iterate over all entries $(s,d,a)$ in $\mathcal{T}$, where $s$ and $d$ are the source- and destination-chiplet-IDs and $a$ is the amount of traffic between them.
For each such entry, we calculate the path ${route}(s,d)$ from $s$ to $d$ using the \textit{routing table} and we sum up all vertex- and edge-weights on the path. 
We compute the average latency over all entries $(s,d,a) \in \mathcal{T}$, weighted by the amount of traffic $a$ between chiplets $s$ and $d$:
\begin{equation*} \label{eq:}
	\small
	L = \frac{\sum \limits_{(s,d,a) \in \mathcal{T}} a \cdot \left ( w(s) + \sum \limits_{\{u,v\} \in \text{route}(s,d)} w(\{u,v\}) + w(v) \right )}{\sum \limits_{(s,d,a) \in \mathcal{T}} a}
\end{equation*}

\subsubsection{Throughput Proxy}
\label{sssec:rc-core-throughput}

We use the same graph representation as for the latency proxy, where we compute two new properties for each edge $\{u,v\} \in E$: the bandwidth $B(\{u,v\})$ and the flow $F(\{u,v\})$. 
The bandwidth $B(\{u,v\})$ depends on the number of bumps that are available to connect the link to its adjacent chiplets $c \in \{u,v\}$ and on the number of non-data wires $N_{ndw}$ that are required for clock or handshaking signals.
Said number of bumps depends on the chiplet area $A_c$, the bump pitch $P_c$, and the fraction $f_{c,\{u,v\}}$ of the chiplet-area available for the bumps of the link $\{u,v\}$. 
The flow $F(\{u,v\})$ is the sum of the traffic in $\mathcal{T}$ that is routed over the edge $\{u,v\}$:
\begin{equation*} \label{eq:}
\vspace{-0.5em}
	\scriptsize
	B(\{u,v\}) = \left \lfloor \frac{A_c \cdot f_{c,\{u,v\}}}{P_c^2} \right \rfloor - N_{ndw}
	\hspace{0.75em} \vline \hspace{0.75em}
	F(\{u,v\}) = \hspace{-1em}\sum \limits_{\substack{(s,d,a) \in \mathcal{T} \\ \{u,v\} \in \text{route}(s,d)}} \hspace{-2em} a \hspace{2em}
\end{equation*}
Based on these two edge properties, we estimate the total throughput $T$ under a given \textit{traffic} $\mathcal{T}$.
Notice that dividing $B(\{u,v\})$ by $F(\{u,v\})$ gives a metric for how much traffic the link $\{u,v\}$ can tolerate expressed as a fraction of the traffic that is currently routed over it, i.e., if this metric is smaller than 1, the link tolerates less traffic than it currently carries and if it is larger than 1, the link can tolerate more traffic.
We estimate how much load the whole \gls{ici} can tolerate expressed as a fraction of the total traffic by taking the minimum of that metric over all edges $\{u,v\} \in E$:
We multiply this metric by the total traffic to get the throughput $T$:
\begin{equation*} \label{eq:}
\vspace{-0.5em}
	\small
	T = \min \limits_{\{u,v\} \in E} \frac{B(\{u,v\})}{F(\{u,v\})} \cdot \sum \limits_{(s,d,a) \in \mathcal{T}} \hspace{-1em}a\hspace{1em}
\end{equation*}
This throughput proxy might look overly pessimistic, however, it turns out to be a good approximation for congestion since as soon as a link is saturated, the congestion starts to quickly spread across the network and also block flows that do not use the saturated link.
In \Cref{sec:eval}, we see that this proxy sometimes over- and sometimes underestimates the throughput.

\subsubsection{Area, Power, and Cost Reports}
\label{sssec:rc-core-overheads}

RapidChiplet reports the total chiplet area, the interposer area, the power consumption, and an estimate for the manufacturing cost.
The chiplet area report straightforwardly sums up the areas of all chiplets and the interposer area is that of the smallest possible rectangle that encloses all chiplets.
These reports are intended for the automated \gls{dse} where the number, size, and placement of chiplets are automatically adjusted based on the design parameters.
The same motivation applies to the power report, which, in addition to summing up the per-chiplet power and the interposer-power, can also incorporate the length-dependent power consumption of links.
The manufacturing cost report uses a simple yield model to estimate the per-chiplet cost, which depends on the chiplet area and on technology-dependent inputs.
These per-chiplet costs are summed up and added to the packaging cost to estimate the total system cost.

\subsection{Cycle-based Simulations using BookSim2}
\label{ssec:rc-bs}

To thoroughly evaluate a small number of promising designs that are identified using RapidChiplet's performance proxies, we support a seamless integration of the cycle-based, flit-level simulator BookSim2 \cite{booksim}. 
The topology, including potentially link-length-dependent latencies, as well as routing tables, traffic patterns, and traces are exported to BookSim2.
We extended BookSim2 with the capabilities to load custom routing tables and traffic patterns, and to simulate network traces, including dependency-tracking (out-of-the-box BookSim2 does not support trace-based simulations).
To perform simulations using BookSim2, the user is required to supply a \textit{booksim\_config} file with simulation parameters.

\vspace{-0.5em}
\subsection{Automated Design Space Exploration}
\label{ssec:rc-dse}

We provide a rich toolbox to automatically explore the design space of \gls{ici}s.
The user can specify ranges of topologies, chiplet counts and sizes, traffic patterns, traces, technology nodes, packaging technologies, latency profiles, and many more parameters.
Our toolchain then iterates over all possible combinations of those parameters, generates the corresponding input files, and runs RapidChiplet.
We proceed by describing our input generation scripts, which the user can easily extend with custom topologies, routing algorithms, placements, etc.

\subsubsection{Chiplets}
\label{sssec:rc-dse-chiplets}

The toolchain creates chiplets with a configurable base area and power, plus a configurable area and power overhead per PHY.
The number of PHYs is determined by the topology.
Our chiplet generator supports different PHY placements (see \Cref{fig:phy-placements}) where the toolchain automatically selects the most suitable PHY placement for a give topology.

\input{fig_phy_placements}

\subsubsection{Placements}
\label{sssec:rc-dse-plac}

RapidChiplet offers two chiplet placements: A 2D grid that is used for most topologies and a hexagonal placement that is used for the HexaMesh \cite{hexamesh} topology and its derivatives.
The placement generator supports any chiplet size and a configurable spacing between chiplets.

\subsubsection{ICI Topologies}
\label{sssec:rc-dse-topo}

RapidChiplet generates the following \gls{ici} topologies for configurable chiplet-counts:
2D Mesh, 
2D Torus, 
Folded 2D Torus \cite{folded-torus},
Flattened Butterfly \cite{flattened-butterfly},
HexaMesh \cite{hexamesh},
HexaTorus,
Folded HexaTorus,
OctaMesh,
OctaTorus,
Folded OctaTorus,
Hypercube \cite{hypercube},
Double Butterfly \cite{double-butterfly},
ButterDonut \cite{butterdonut},
ClusCross \cite{cluscross},
Kite \cite{kite}, and
SID-Mesh \cite{sid-mesh}.

\subsubsection{Routing Tables}
\label{sssec:rc-dse-routing}

Our routing table generator creates a routing table for each chiplet and on-interposer router. 
Notice that since RapidChiplet supports arbitrary topologies, routing algorithms for mesh network such as XY routing are not applicable.
We support two different, deadlock-free routing algorithms for arbitrary topologies, which are both based on Dijsktra's algorithm \cite{dijkstra}.
The first one routes each packet along the shortest path and in the case of multiple shortest paths, it selects the next hop with the lowest ID to avoid deadlocks.
This is the same deadlock-avoidance strategy as in BookSim2's routing algorithm for arbitrary topologies, however, this strategy fails to exploit path diversity.
To alleviate this issue, we offer a second, randomized routing algorithm that uses the turn model \cite{turn-model}, the simple cycle braking algorithm \cite{cycle-breaking}, and a dual graph construction \cite{dual-graph} for deadlock-free, shortest-path routing.

\subsubsection{Traffic}
\label{sssec:rc-dse-traffic}

We offer generators for the synthetic \textit{random-uniform}, \textit{transpose}, \textit{permutation}, and \textit{hotspot} traffic patterns.

\subsection{Exporting Traffic Traces using Netrace}
\label{ssec:rc-netrace}

We provide a script to export traces from the Netraces v1.0 trace collection \cite{netraces} using Netrace \cite{netrace,netrace-report}, and to parse them into the RapidChiplet format. 
These traces contain traffic between the L1 cache, L2 cache, and main memory, and they use source and destination addresses in the range $0$ to $63$.

\vspace{-0.5em}
\subsection{Visualization}
\label{ssec:rc-vis}

RapidChiplet can visualize its inputs (see \Cref{fig:visualization} left) which is especially helpful when creating placements and topologies by hand, when extending the generator-scripts with new placements or topologies, or to analyze the solution that an automated \gls{dse} or optimization algorithm found.
We also provide some functions to visualize results, e.g, latency-vs-load plots based on BookSim-simulations (see \Cref{fig:visualization} right).

\input{fig_visualization}

%% file: fig_overview.tex
\begin{figure*}[h]
\centering
\captionsetup{justification=centering}
\includegraphics[width=1.0\textwidth]{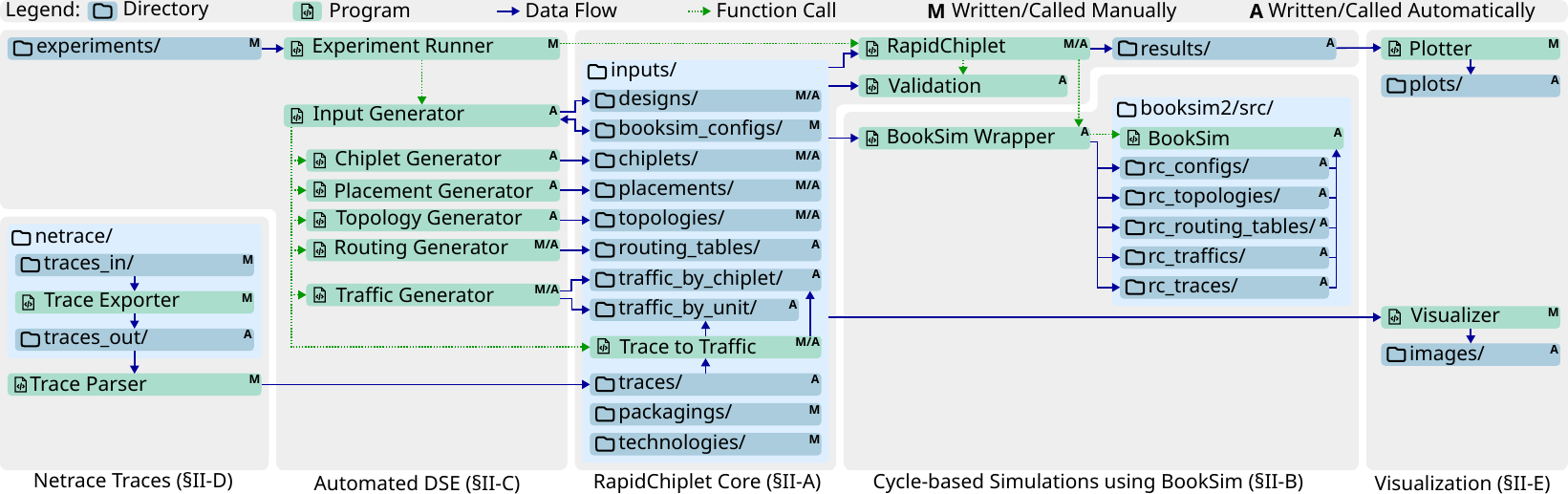}
\caption{\textbf{(\textsection \ref{sec:rc}) RapidChiplet architecture overview}.}
\vspace{-1em}
\label{fig:overview}
\end{figure*}

%% file: fig_inputs.tex
\begin{figure*}[b]
\centering
\captionsetup{justification=centering}
\includegraphics[width=1.0\textwidth]{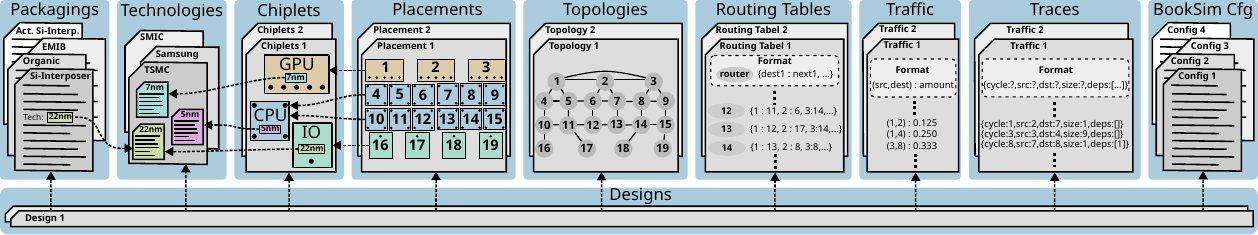}
\caption{\textbf{(\textsection \ref{sssec:rc-core-inputs}) Overview of RapidChiplet input files} and how they reference each other.}
\label{fig:inputs}
\end{figure*}

%% file: fig_phy_placements.tex
\begin{figure}[h]
\vspace{-1.0em}
\centering
\captionsetup{justification=centering}
\includegraphics[width=1.0\columnwidth]{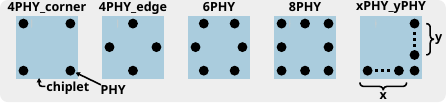}
\vspace{-2.5em}
\caption{\textbf{(\textsection \ref{sssec:rc-dse-chiplets}) Available PHY placements}}
\label{fig:phy-placements}
\end{figure}
\vspace{-2.0em}

%% file: fig_visualization.tex
\begin{figure}[h]
\vspace{-0.25cm}
\centering
\captionsetup{justification=centering}
\includegraphics[width=1.0\columnwidth]{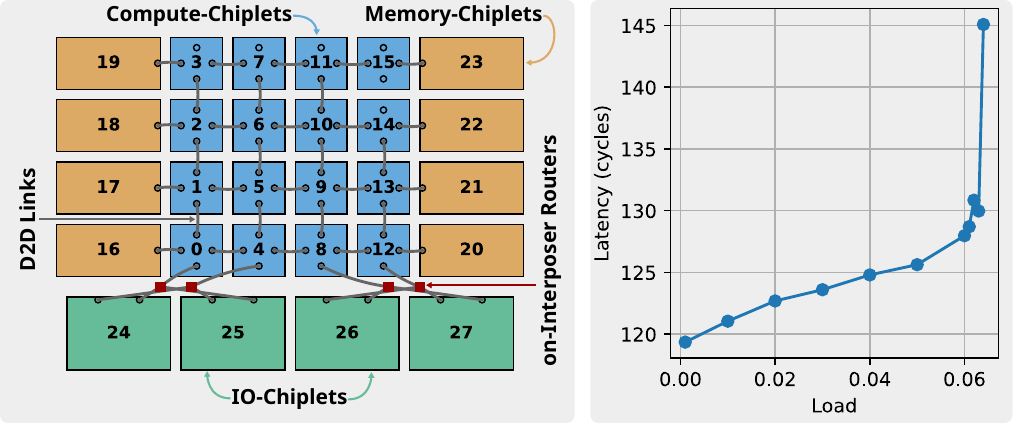}
\vspace{-0.5cm}
\caption{\textbf{(\textsection \ref{ssec:rc-vis}) Visualized design (left); results-plot (right)}.}
\label{fig:visualization}
\vspace{-0.5cm}
\end{figure}

%% file: 03_evaluation.tex
\section{Evaluation}
\label{sec:eval}

\input{fig_evaluation}

To evaluate the accuracy and speed of our latency and throughput proxies, we compare them against cycle-based flit-level simulations in BookSim2 \cite{booksim}.

\subsection{Experiment Setup and Methodology}
\label{ssec:eval-setup}

We evaluate RapidChiplet using the Mesh, Torus, Folded Torus \cite{folded-torus}, and SID-Mesh \cite{sid-mesh} topologies, the random-uniform, transpose, permutation, and hotspot\footnote{Four hotspot nodes and $50\%$ of the traffic is directed towards these hotspots.} traffic patterns, and chiplet counts ranging from $9$ up to $100$.
We set the latency of chiplets and PHYs to $3$ and $12$ cycles respectively, and we set the link-latency to $0.25$ cycles/mm.
Furthermore, we use square chiplets with a base-area of $74$mm$^2$ and an area overhead of $0.85$mm$^2$ per PHY.
BookSim2 models input-queued routers with a pipeline of routing-, VC allocation-, switch allocation-, and crossbar traversal-stages.
We use four virtual channels with 16-flit buffers each, and we adjust the number of simulated cycles to the number of chiplets.
While a single BookSim-simulation is sufficient to get the zero-load latency, we need to run multiple simulations with increasing injections rates to determine the saturation throughput.
We first increase the injection rate in $10\%$-steps and once saturation is reached, we go back to the last stable injection rate and proceed with $1\%$-steps and finally with $0.1\%$-steps.
E.g., determining a saturation throughput of $12.3\%$ requires $9$ simulations with the injection rates $10\%,20\%,11\%,12\%,13\%,12.1\%,12.2\%,12.3\%,12.4\%$.
The full list of parameters can be found in our repository\footnote{ \ifnb https://github.com/spcl/rapidchiplet \fi }.
All experiments are performed on a laptop PC with an Intel Core i7-1165G7 CPU running Arch Linux with kernel version 6.10.6.

\subsection{Experimental Results}
\label{ssec:eval-results}

\Cref{fig:evaluation} shows the deviation of RapidChiplet's latency and throughput proxies from the BookSim2 simulation results (relative error) and RapidChiplet's speedup over BookSim2.

\subsubsection{Latency Proxy}
\label{sssec:eval-results-latency}

On average, our latency proxy is $1075\times$ faster than simulations in BookSim2, which comes at the cost of an average error of $2.57\%$.
The error mainly depends on the traffic pattern: For random-uniform and transpose traffic, the average error is lower than $0.3\%$, for hotspot traffic, it is $1.66\%$ and for permutation traffic, it is $8.11\%$.
The speedup depends on the number of chiplets and the traffic pattern.
For the random-uniform and hotspot traffic patterns, the speedup grows linearly with the number of chiplets, while for the transpose and permutation traffic patterns, the speedup grows quadratically with the number of chiplets.
This behavior can be explained by analyzing the runtime\footnote{Runtime results are omitted due to space constraints. They can be found in our repository:  \ifnb https://github.com/spcl/rapidchiplet \fi } of RapidChiplet and BookSim2.
The runtime of BookSim2 grows quadratically with the number of chiplets, ranging from about $0.05$ s for 9 chiplets to about $13$ s for 100 chiplets (independently of the traffic pattern).
The runtime of RapidChiplet grows linearly in the number of communicating chiplet-pairs, which grows quadratically in the number of chiplets for the random-uniform and hotspot traffic patterns and linearly for the transpose and permutation traffic patterns, hence, the difference in speedup between the traffic patterns.
The simulated \gls{ici} topology does not seem to affect the error or speedup of RapidChiplet's latency proxy.

\subsubsection{Throughput Proxy}
\label{sssec:eval-results-throughput}

Our throughput proxy achieves an average speedup of $69$,$079\times$ over BookSim2 with an average error of $25.12\%$.
The runtime of RapidChiplet's throughput proxy follows the same trends as its latency, proxy, i.e., it grows linearly with the number of chiplets for the transpose and permutation traffic patterns and quadratically for the random-uniform and hotspot traffic patterns.
The combined runtime of the multiple BookSim2-simulations needed to identify the saturation throughput grows more or less linearly with the number of chiplets, ranging from about $10$ s for 9 chiplets to about $300$ s for 100 chiplets. 
Therefore, the speedup of our throughput proxy declines with an increasing chiplet count for the random-uniform and hotspot traffic patterns, while it stays more or less constant for transpose and permutation traffic.

Another interesting observation is that the average speedup of the throughput proxy is $64\times$ higher than that of the latency proxy.
The reasons for this are twofold:
Firstly, as explained in \Cref{ssec:eval-setup}, we need to run multiple simulations with increasing injection rates to determine the saturation throughput.
Secondly, and most importantly, simulations close to saturation (as required to determine the saturation throughput) are significantly more time-consuming than simulations at low injection rates (as required to determine the zero-load latency).

The huge speedup of RapidChiplet's throughput proxy comes at the cost of a rather high approximation error.
These results fortify our approach of using performance proxies for a large-scale \gls{dse} where simulations are not feasible, and using simulations to accurately evaluate a set of selected design points.

%% file: fig_evaluation.tex
\begin{figure*}[h!]
\centering
\captionsetup{justification=centering}
\includegraphics[width=0.965\textwidth]{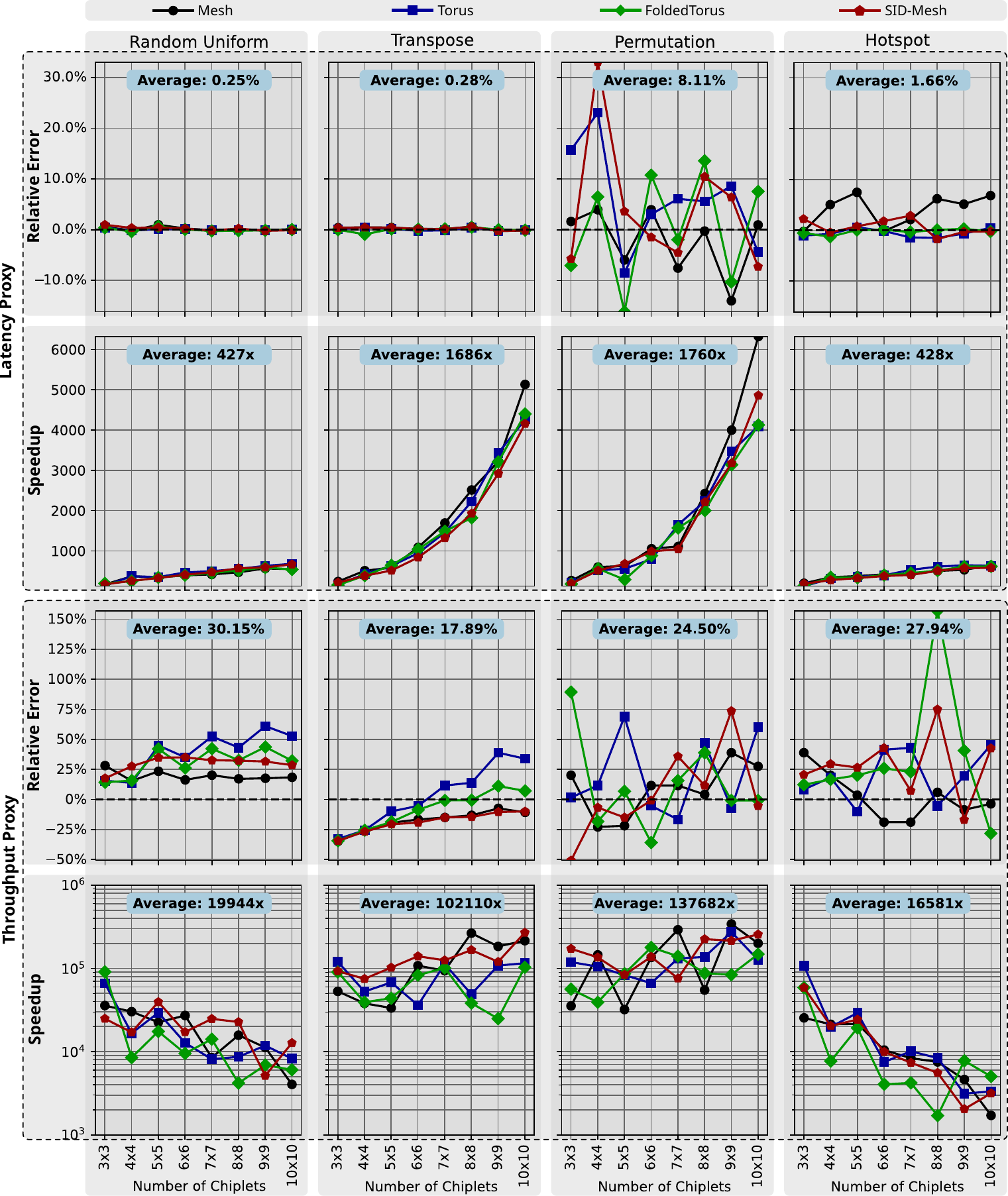}
\caption{\textbf{(\textsection \ref{ssec:eval-results}) Evaluation of RapidChiplet's latency and throughput proxies} compared to simulations in BookSim2.}
\label{fig:evaluation}
\vspace{2em}
\end{figure*}

%% file: 04_case_study.tex
\section{Case Study}
\label{sec:cs}

To showcase the possibilities that RapidChiplet unveils, we perform a case study in which we analyze the performance of the \gls{shg} \gls{noc} topology, if used as an \gls{ici}.
\gls{shg} \cite{shg} is a parametrizable \gls{noc} topology that spans the design space between a 2D mesh (low area but low performance) and a flattened butterfly (high area but high performance). 
For a grid of $R \times C$ chiplets, there are $2^{R+C-4}$ different parametrizations.
Since for large designs, thousands of possible \gls{shg} parametrizations exist, the authors propose a manually driven selection strategy that does not explore the whole design space.
We exploit the unprecedented speed of RapidChiplet to evaluate the $65'536$ possible parametrizations of \gls{shg} for a $10\times10$ chiplet grid, which takes less than half a day on a consumer-grade laptop.

\subsection{Experiment Setup}
\label{ssec:cs-setup}

We use a $10\times10$ grid of homogeneous chiplets with the same specifications as in our evaluation (see \Cref{ssec:eval-setup}) and we estimate the latency and throughput of each \gls{shg} parametrization under the random-uniform traffic pattern.

\subsection{Experiment Results}
\label{ssec:cs-results}

\Cref{fig:case-study} shows the results of the case study.
Each point in the plot corresponds to one of the $65'536$ \gls{shg} parametrizations, with the 2D mesh and flattened butterfly topologies highlighted.
The color of the points represents the area overhead, ranging from $0\%$ for the 2D mesh to $16\%$ for the flattened butterfly.
Lines show latency-throughput Pareto-fronts under different area-constraints (labeled with the area overhead vs. a 2D mesh).

\input{fig_case_study}

We observe that a high area overhead is a necessary condition for both low latency and high throughput.
While a larger area budget almost seems to be a sufficient condition for low latency, this does not hold for the throughput as there are many points which, despite having a high area overhead, suffer from a low throughput.
We conclude that in order to achieve a high throughput under a given area-budget, finding a good parametrization is crucial.
The best way to find this parametrization is through exhaustive search, which is enabled by RapidChiplet's fast latency and throughput proxies.

%% file: fig_case_study.tex
\begin{figure}[h]
\vspace{-1.0em}
\centering
\captionsetup{justification=centering}
\includegraphics[width=1.0\columnwidth]{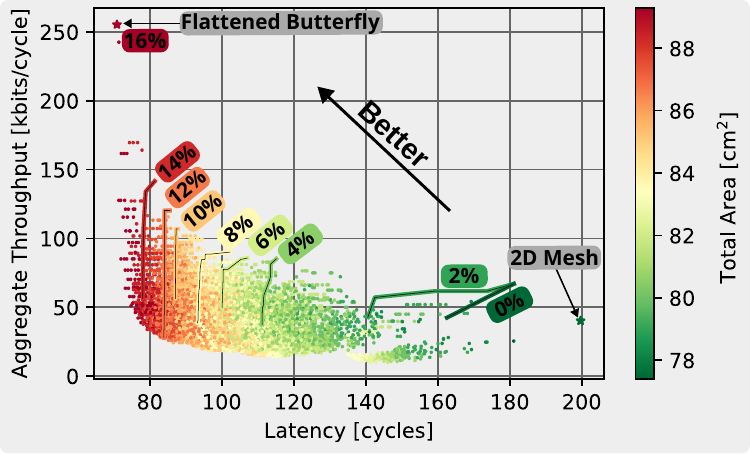}
\caption{\textbf{(\textsection \ref{ssec:cs-results}) Design space exploration of the SHG topology}. Points are configurations; lines are latency-\\throughput pareto-frontiers for different area-overheads.}
\label{fig:case-study}
\vspace{-0.5em}
\end{figure}

%% file: 05_related_work.tex
\section{Related Work}
\label{sec:rl}

Early-stage latency and throughput predictions for ICIs are often performed using cycle-based simulators such as BookSim2 \cite{booksim}, Noxim \cite{noxim}, Nostrum \cite{nostrum} or Garnet \cite{garnet}.
While these simulators were originally conceived to simulate \gls{noc} of monolithic chips, methodologies \cite{chip-sim-method} to use them for simulations of 2.5D stacked chips have been proposed.
As we have shown, such simulations take orders of magnitude longer than our novel, high-level latency and throughput proxies.

There exist numerous \gls{dse}-tools for other metrics, such as the Orion 2.0 \cite{orion} power and area model, the ChipletActuary \cite{chiplet-actuary} cost model, or the HotSpot \cite{hotspot} thermal simulator. RapidChiplet focuses on the latency and throughput of the \gls{ici} and only provides very high-level power, area, and cost estimates to assess the influence of the \gls{ici} onto those metrics.

Numerous works perform a \gls{dse} of chiplet architectures.
Pal et al. \cite{chiplet-dse-1} present a \gls{dse} for chiplet systems, focusing on finding a minimum set of chiplets as building blocks for a processor-family.
Chiplet-Gym \cite{chiplet-gym} is a reinforcement learning-based framework to explore the design space of chiplet-based accelerators for artificial intelligence workloads.
Kim et al. \cite{chiplet-dse-2} present a design flow that covers the architecture, circuit and packaging design phases of a 2.5D integrated circuit.
Our work focusses on the \gls{dse} of the \gls{ici} specifically, rather than the overall architecture.

%% file: 06_conclusion.tex
\vspace{-0.5em}
\section{Conclusion}
\label{sec:conclusion}

\glsreset{ici}

The design space for \gls{icis} is huge, as there are countless options for packaging, chiplet placement, \gls{d2d} link implementation, \gls{ici} topology and many more, which all influence the performance and cost of the \gls{ici}.
Exploring this large design space requires a fast framework capable of capturing the complex interplay between these design choices.

As we are not aware of an existing toolchain satisfying these requirements, we propose RapidChiplet, a fast and comprehensive toolchain that provides high-level latency and throughput proxies for \gls{icis} based on a multitude of configurable parameters.
Compared to cycle-based simulations, our latency proxy achieves an average speedup of $1,075\times$ at an average error of $2.57\%$.
Our throughput proxy achieves a higher average speedup of $69,079\times$ at a larger average error of $25.12\%$.

RapidChiplet can be used for \gls{dse}s where hundreds of thousands of design points are evaluated, or as a cost function to optimization algorithms or machine learning models.
By providing RapidChiplet with its novel, high-level latency and throughput proxies, we facilitate \gls{dse} of \gls{icis}, sparking future advancements in this active research area.

%% file: 07_acknowledgements.tex
\vspace{-0.5em}
\section*{Acknowledgements}
\label{sec:ack}
\vspace{-0.125em}

\iftrue
This work was supported by the ETH Future Computing Laboratory (EFCL), financed by a donation from Huawei Technologies.
It also received funding from the European Research Council
\raisebox{-0.25em}{\includegraphics[height=1em]{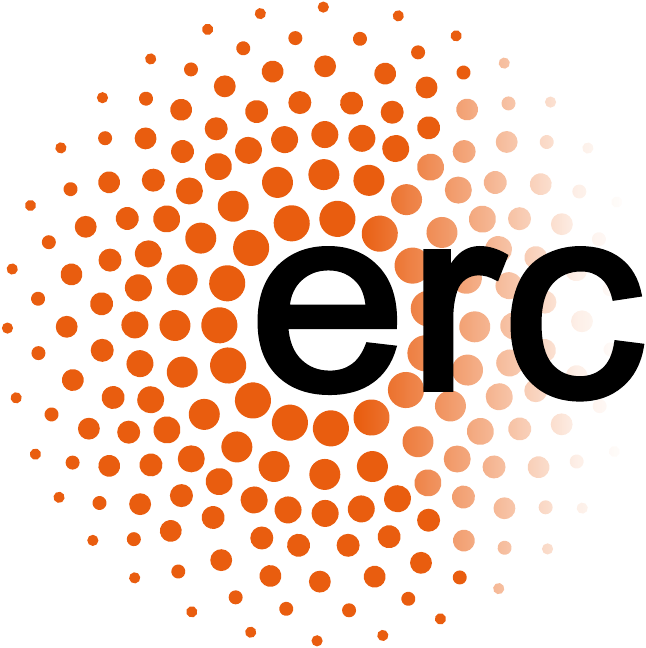}} (Project PSAP,
No.~101002047), the European Union's HE research and innovation programme (Project GLACIATION, No.~101070141), and the European PILOT project which has received funding from the European High-Performance Computing Joint Undertaking (No.~101034126).
\fi

